# INDIVIDUAL FACTORS THAT INFLUENCE EFFORT AND CONSTRIBUTIONS ON WIKIPEDIA

**authors:** Luiz F. Pinto, Carlos Denner dos SANTOS (carlosdenner@unb.br), Silvia Onoyama

**ABSTRACT**

In this work, we aim to analyze how attitude, self-efficacy, and altruism influence effort and active contributions on Wikipedia. We propose a new conceptual model based on the theory of planned behavior and findings from the literature on online communities. This model differs from other models that have been previously proposed by considering altruism in its various facets – by identification, reciprocity, and reputation – and by treating effort as a factor prior to performance results, which is measured in terms of active contributions, according to the organizational literature. To fulfill the study's specific objectives, Wikipedia surveyed community members and collected secondary data. After excluding outliers, we obtained a final sample with 212 participants. We applied exploratory factor analysis and structural equation modeling, which resulted in a model with satisfactory fit indices. The results indicate that effort influences active contributions, and attitude, altruism by reputation, and altruism by identification influence effort. None of the proposed factors are directly related to active contributions. Experience directly influences self-efficacy while it positively moderates the relation between effort and active contributions. Finally, we present these conclusions' implications for the literature and Wikipedia and suggestions for future studies.

**Key words: Online communities, effort, contributions, performance results, theory of planned behavior**

# 1. INTRODUCTION

With the rise of the Internet, collaborative platforms, social networks, blogs, and online communities emerged and redefined the relationship between businesses and consumers and amongst people themselves. Communities with various purposes were created, and people with common interests may share information and knowledge much more easily today.

In this scenario, we have broken physical barriers since the online community can be defined as a collective of people whose main objective is the social well-being of its members. It is where participants share common interests, experiences, or convictions and interact with each other via the Internet (Sproull, 2003). Examples of online communities include electronic databases of knowledge, discussion forums, and Wiki platforms, collaborative encyclopedias whose main purpose is to spread the knowledge generated by an online community to as many people as possible.

Firms may play several roles in online communities, such as observe and gather information, host or sponsor communities (create and manage web sites and advertisements), provide content to communities (such as music, information, or entertainment), and participate as members in direct relationships with other participants (Miller, Fabian, & Lins, 2009). As contributions are voluntary in online communities, participants' efforts and contributions vary enormously. While some contribute little or nothing, others contribute a considerable part or most of the content, (Anthony, Smith, & Williamson, 2007; Levine & Prietula, 2013; Ren, Kraut, & Kiesler, 2007).

From a purely rational perspective, collaborating with an online community does not appear to make sense since members spend time and resources without ever expecting some individual gain in return (Anthony et al., 2007). In recent years, studies have explored the factors that motivate such behavior, and most base themselves on the theory of planned behavior (Cho, Chen, & Chung, 2010; Lin, 2006; Tsai & Bagozzi, 2014). They analyze the relations among variables such as attitude, perceived behavioral control – understood as self-efficacy, and subjective norms. Other works, in turn, investigate the behavior behind contribution and the factors that determine such behavior (Chang & Chuang, 2011; Park, Oh, & Kang, 2012; Wasko & Faraj, 2005).

The issue of work performance results, which we consider in terms of active contributions to the online community, is fundamental to the organizational literature. Recognizing their importance, Ajzen (2011) proposed a new perspective to explain work performance results based on employees' efforts through the original components of the theory of planned behavior.

Internationally, Cho et al. (2010), Park et al. (2012), and Xu and Li (2015) have proposed three models that aim to test the relation between specific factors and contributions on Wikipedia involving the original components of the theory of planned behavior and other differential variables, such as altruism. Therefore, we propose a new explanatory model for the phenomenon. The theory of planned behavior provides more solid hypotheses since it can satisfactorily predict the antecedents of certain individual behavior in a wide range of situations (Armitage & Conner, 2001), and the information systems literature refers to it extensively.

This model differs from previous models in that it considers effort to be a factor prior to performance results, altruism in three dimensions, and the moderating role of experience in the result of the contributions. Moreover, none of the three models specifically measured active contributions, as we propose herein.

Considering these issues, for this paper, we studied the Wikipedia online community to try to answer the following research question: How do attitude, self-efficacy, and altruism influence effort and active contributions in an online community? Here, we discuss the theoretical basis for the hypotheses and research model, our methodological procedures, analysis, results, and final considerations.

## 2. THEORETICAL FRAMEWORK

We organized the theoretical framework for this article to provide a better understanding of the subject we explore herein and to demonstrate how we constructed the theoretical model based on the theory of planned behavior.

### 2.1. Constructing the Theoretical Model Based on the Theory of Planned Behavior

#### 2.1.1. Effort

Effort can be understood as the amount of time and energy that an employee invests in their work (Campbell & Pritchard, 1976). As opposed to performance results, which are the results expected from the work, effort refers to behavior that enables such results (Christen, Iyer, and Soberman, 2006).

According to expectancy theory, developed by Vroom (1964), the effort a worker exerts (behavior) is determined by their expectation that an increased effort will lead to a certain performance level (result), multiplied by the subjective value given to this level of performance results. Thus, individuals who believe that an elevated level of effort will provide favorable performance results that will offer them positive consequences will put in more effort (Vroom, Porter, & Lawler, 2005).

In more traditional firms, the effort a worker puts forth is oftentimes related to how much they expect to gain financially from the additional effort (Vroom et al, 2005). In online communities, where identification with the group tends to be strong, individuals exert effort to make contributions in favor of the collective whole even though they do not receive organizational compensation (Kankanhalli, Tan, & Wei, 2005)

Initially, we may possibly conclude that effort and the contributions themselves represent the same phenomenon in an online community. However, effort is related more to the time and resources a member spends, while performance results, or contributions, may be conceived as the result of the member's effort.

Although the literature on online communities widely ignores the role of individual effort in determining members' level of contribution (Chang & Chuang, 2011; Kankanhalli et al., 2005; Xu & Li, 2015), we hypothesize the following based on findings from the organizational literature:

> *Hypothesis 1. When an individual exerts more effort, they will make more active contributions.*

### 2.1.2. The relation among experience, effort, contributions, and self-efficacy

**The events that an individual has lived and that are related to the performance results they have attained from executing certain work refers to experience in the organizational context (Quinones, Ford, & Teachout, 1995). In online communities, members' prior experience might influence performance results as older members tend to**

contribute more than new members (Marchi, Gianchetti, & Gennaro, 2011; Ransbotham & Kane, 2011). This is because older members are more engaged with the communities' dynamics and rules and have a better sense of what is expected of them in terms of duties and contributions and face less ambiguity than newcomers in the decision to contribute (Tsai & Bagozzi, 2014).

Upon studying an online image-sharing community, Nov, Naaman, and Ye (2010) found that the number of years registered in the community is directly related to an increase in participation in terms of tags, contacts, and groups and negatively correlated to the number of shared photos. Nevertheless, while investigating an online discussion forum, Bateman, Gray, and Butler (2011) found that members with more years of experience are more susceptible to reading the questions and less susceptible to answering them.

Considering the low correlation between experience and performance results and the ambiguous results indicated by the literature on online communities, it is possible to assume that the relation between the two factors may not be direct. In the model proposed herein, we considered individual effort to determine the amount of contributions, with experience moderating such a relation.

Additionally, the literature commonly points to the significance of the relation between experience and self-efficacy. In other words, the more experience an individual has in a certain profession or task, the more confident they will be in their skills and abilities to execute them. This relation has already been solidified in several contexts, such as teachers' self-efficacy in the classroom (Wolters & Daugherty, 2007) and self-efficacy in computer (Cassidy & Eachus, 2002) and Internet use (Eastin & Larose, 2000).

Thus, we propose the following:

***Hypothesis 2a. More experience means a stronger relation between effort and active contributions.***

*Hypothesis 2b. More experience means greater self-efficacy.*

### 2.1.3. The relation among attitude, effort, and contributions

Attitude refers to the degree in which an individual favorably or unfavorably assesses certain behavior (Ajzen, 1991; Ajzen, 2011). According to the theory of planned behavior, behavioral beliefs directly precede attitude and produce either a more favorable or unfavorable attitude with respect to a given behavior (Ajzen, 2002).

In crowdsourcing initiatives, virtual environments where geographically dispersed firms and individuals interact with each other, attitude seems to be especially relevant to determining participants' intention of contribution. Pinto and Júnior (2015) analyzed two crowdsourcing platforms and found that attitude responds to 72 percent of the variation of the collaborators' intention to contribute.

Similarly, the literature on online communities recognizes that the more a member believes that the act of contributing is pleasant and favorable to them, the greater their chances of contribution (Cho et al., 2010; Lin, 2006; Tsai & Bagozzi, 2014). Tsai and Bagozzi (2014) found that a high attitude level is directly related to a greater willingness to contribute content. In turn, Cho et al. (2010) and Lin (2006) obtained the result that attitude is directly related to the intention to contribute.

According to the theory of planned behavior, the more an individual believes that the consequences of their work efforts are positive, the more favorable their attitude toward work will be. Consequently, they will put forth more effort (Ajzen, 2011). Therefore, we hypothesize the following:

>   *Hypothesis 3a. A more favorable attitude means greater effort.*

>   *Hypothesis 3b. A more favorable attitude means more active contributions.*

### 2.1.4. Perceived behavioral control (self-efficacy) and its relation to effort and contributions

Perceived behavioral control may be understood as an individual's perception regarding the difficulty or facility to act a certain way (Ajzen, 1991; Conner & Armitage, 1998). Accordingly, when an individual becomes more aware of the control they have over the means necessary to executing a certain task, they feel more confident in their capability of executing it (Ajzen, 1991). Based on the theory of planned behavior, perceived behavioral control comes from a series of individual beliefs over the presence of factors that would inhibit or support the manifestation of a given behavior, the beliefs in control (Ajzen, 2002).

The notion of self-efficacy is intimately related to the notion of perceived behavioral control, although most studies on the theory of planned behavior clearly distinguish the two concepts (Ajzen, 2002; Armitage & Conner, 2001; Pavlou & Fygenson, 2006). Generally, perceived behavioral control is considered to be a factor of the highest order containing both self-efficacy and controllability (Ajzen, 2002).

In this context, self-efficacy is understood to be the individual's perception of the degree in which they possess the capabilities and abilities needed to carry out a given task (Bandura, 1982). Controllability refers to personal judgments on the availability of resources and opportunities to exhibit certain behavior (Pavlou & Fygenson, 2006).

Fishbein and Cappella (2006) developed an integrative model of human behavior where perceived behavioral control and self-efficacy are considered equal concepts. In the context of online communities, this consideration is particularly relevant as their members have ample access to the means to contribute (Levine & Prietula, 2013). Only self-efficacy is a relevant factor, while controllability appears not to be so.

According to Bandura (1982), efficacy expectations in an activity's execution will determine how much effort an individual is willing to put forth and how determined they will be to persist even when faced with challenges. Therefore, people who believe they can perform a task satisfactorily will have better performance results than those who believe they will fail to perform (Gist & Mitchell, 1992).

Upon analyzing 158 articles, Judge, Jackson, Shaw, Scott, and Rich (2007) also confirmed the importance of self-efficacy only in certain contexts and in interaction with specific variables. For example, the authors found that self-efficacy predicts performance results for simple tasks but not for intermediate and complex tasks. Similarly, it can only predict results for specific tasks but not overall work performance results.

In the literature on online communities, self-efficacy is commonly considered to be an influential factor in the participants' intention to contribute and effective contribution (Cho et al., 2010; Hsu, Ju, Yen, & Chang, 2007; Kankanhalli et al., 2005; Park et al., 2012; Ray, Kim, & Morris, 2014). In this context, authors analyze electronic databases of knowledge (Kankanhalli et al., 2005), Wiki platforms (Cho et al., 2010; Park et al., 2012), and discussion forums (Hsu et al., 2007; Ray et al, 2014).

Kankanhalli et al. (2005) found that the self-efficacy of knowledge – that is, how much individuals believe that their knowledge may be useful to the community – is directly related to the use of an electronic database. Self-efficacy of knowledge is also important in favoring engagement in a discussion forum (Ray et al., 2014) and the intention to contribute content to a collaborative encyclopedia (Park et al., 2012).

Members with greater perceived self-efficacy make regular, quality contributions to an online community to the extent where they perceive that their participation is crucial to providing it with content and that contribution promotes their self-image as effective people in general (Kollock, 1999).

Considering the relevance of self-efficacy in determining individuals' efforts to execute tasks (Ajzen, 2011) and findings in the literature on online communities concerning the importance of self-efficacy in members' contributions, we propose the following hypotheses:

*Hypothesis 4a. More self-efficacy means greater effort.*

*Hypothesis 4b. More self-efficacy means more active contributions.*

### *2.1.5. Subjective norms in the context of online communities*

Subjective norms refer to the social pressures that an individual perceives upon presenting a certain behavior or not (Ajzen, 1991). This factor arises directly from normative beliefs, which are beliefs regarding other people's expectations (Ajzen, 2002).

Upon conducting a meta-analysis with 185 articles that applied the theory of planned behavior as a theoretical approach, Armitage and Conner (2001) found that subjective norms have significantly less predictive power than the theory's other two antecedents, attitude and perceived behavioral control, when determining intention. Similarly, research on online communities has systematically and even empirically indicated subjective norms' inefficacy as a predictive factor of the intention to contribute (Cho et al., 2010; Park et al., 2012; Tsai & Bagozzi, 2014). To ex-plain this result, online communities are distributed virtual environments where face-to-face contact rarely occurs, thus hampering other people's influence on the individual's behavior, as occurs in organizations, for example (Cho et al., 2010).

Considering subjective norms' weak predictive power related to the intention to contribute, including in the context of online communities, we considered this factor not to exert any influence over the participant's effort to contribute effectively.

### 2.1.6. The relation among altruism, effort, and contributions

According to the evolutionary theory, altruism is the behavior that reduces an individual's aptitude while others' aptitudes improve. If the altruist's total contribution to the others' aptitude is greater than the aptitude lost they have lost, altruism will increase the group's chances of survival in a competitive environment (Simon, 1993).

More specifically, in the context of online communities, it is important to study altruism's influence on members' contribution behavior mainly because the individuals involved rarely receive some kind of monetary compensation for their participation in the group (Anthony et al., 2007). Furthermore, many members report that they act in favor of the community as a whole regardless of their own personal interests.

Altruism may be conceived in terms of altruism by identification, by reciprocity, and by reputation (Fehr & Fishbacher, 2003). In the first case, individuals are willing to help those around them or related to them without expecting anything in return (Rose-Ackerman, 1996). In the second, individuals help others while considering the possibility of receiving some form of compensation in the future or acting in response to another's altruistic action (Humphrey, 1997).

In the last instance, people act because they are motivated by the desire to receive recognition from others for their altruistic behavior (Fehr & Fishbacher, 2003).

In an online community, members generally contribute without expecting any kind of direct retribution (Kollock, 1999; Wasko & Faraj, 2005). In this context, participants add content mainly because they wish to repay the efforts made by other collaborators to provide knowledge to the group (Cho et al., 2010). Given the above, reciprocity may positively influence the decision to contribute to an online community (Chang & Chuang, 2011; Cho et al., 2010; Kankanhalli et al., 2005).

Altruistic behavior may also be motivated by the desire to increase reputation. A willingness to help others, active participation, and contributing quality content may improve a member's reputation before their peers in an online community (Kollock, 1999). Therefore, upon noticing that their names are ranked among the most significant contributors, participants feel honored and pleased that their efforts have been recognized (Cho et al. 2010). Consequently, they become more inclined to contribute (Kankanhalli et al., 2005; Xu & Li, 2015; Wasko & Faraj, 2005).

Thus, we pose the following hypotheses:

*Hypothesis 5a. More altruism by identification means more effort.*

*Hypothesis 5b. More altruism by identification means more active contributions.*

*Hypothesis 6a. More altruism by reciprocity means more effort.*

*Hypothesis 6b. More altruism by reciprocity means more active contributions.*

*Hypothesis 7a. More altruism by reputation means more effort.*

*Hypothesis 7b. More altruism by reputation means more active contributions.*

Considering these hypotheses, we propose a new conceptual model, as shown in Figure 1:

---------------------------------------

Insert Figure 1 about here

---------------------------------------

## 3. METHOD

### 3.1. Type and General Description of the Study

**This work is a transversal study. We applied a survey and gathered secondary data.**

### 3.2. Unit of Analysis

**This study aims to examine the Wikipedia online community. We chose this community because several articles in the literature made the same choice (Park et al., 2012; Xu & Li, 2015; Zhang & Wang, 2012). This fact leads us to believe that Wikipedia is an appropriate platform for studying the factors that influence contributions from Wiki platform members. Moreover, it has a significant number of contributors, members with highly collaborative profiles, and an internal group exclusively dedicated to education and academic research, thereby facilitating the collection of questionnaires.**

### 3.3. Operating the Variables

With the exception of the effort construct, we measured the data for the constructs included in the research instrument with a five-point Likert scale. We applied an online questionnaire with 34 items, and the first four items are descriptive of the sample.

**To form the attitude construct, we chose one item from the study conducted by Sai,**

Lim, Leung, Lee, Huang, and Benbasat (2009) and three from the study by Johnston and Warkentin (2010). For self-efficacy, two items come from research by Cho et al. (2010), which also concerns contribution on Wikipedia, two items from the study by Perrewé, Zellars, Ferris, Rossi, Kacmar, and Ralston (2004), and two items from the article written by Kankanhalli et al. (2005). Regarding altruism by identification, we took three items from the article by Boivie, Lange, McDonald, and Westphal (2011) and the other three from Kankanhalli et al. (2005). We found the most appropriate items for the identification construct in articles by Cho et al. (2010); Flynn, Reagans, Amanatullah, and Ames (2006); and Edwards, Cable, Williamson, Lambert, and Shipp (2006), with three items each, respectively. As for altruism by reciprocity, we chose one item from each study by Xu & Li (2014), Cho et al. (2010), and Hofmann and Morgeson (1999).

For the effort construct, we utilized a scale of three items proposed by Rasch and Tosi (1992). This study aims to assess the factors that influence software developers' performance results based on the expectancy theory by Vroom (1964). The authors developed a scale in which effort is measured by multiplying valence, the importance an individual gives to compensation, and expectation, the belief that increased effort will lead to increased performance results. In this study, valence is represented by the perceived allure of being a member with a high contribution level, while expectation is represented by the probability in which the individual perceives that they are a high-impact contributor, multiplied by their current effort.

After defining the final version of the questionnaire, we submitted it for evaluation by judges. Three judges were academic researchers in related topics while the other three were Wikipedia contributors, and one of the academic researchers was also a contributor.

The results we attained served as input for correcting, improving, adding, and excluding items.

We researched documents to obtain data concerning experience, active contributions, and members' user profiles. We measured experience by the number of days the user had been registered in the platform from the day of their first edition until January 25, 2017, and we considered the level of contribution to be the user's total active editions. Thus, it was possible to arrive at a measure that encompasses both quantity and quality (active editions refer to content that has not been eliminated by another user based on criteria such as inappropriateness, bad quality, among others). We also recorded information regarding the user's profile in the platform, that is, if they hold any kind of administrative position or if they are just a common user

### 3.4. Sample

The study's sample includes members registered in the Portuguese version of Wikipedia (http://pt.wikipedia.org). To reach this audience, we sent questionnaires to Wikimedia Brasil's e-mail lists, made an announcement in Wikipedia's notice section, and sent private messages to members through the platform itself.

The process for distributing the questionnaires and collecting data lasted from December 20, 2016 to January 20, 2017, and resulted in the participation of 225 respondents. After verifying the validity of the user informed on the questionnaire (information essential to retrieving secondary data), we reduced this number to 214 respondents. As we chose a sample that does not include all participants of the Portuguese version of Wikipedia and is not random, it is a non-probability sampling by convenience.

**3.5. Data Analysis**

We tested our study's hypotheses with structural equation modeling through the maximum likelihood method. To statistically analyze the data, we used the IBM SPSS version 22 and AMOS version 22 software.

# 4. RESULTS

**4.1. Sample Description**

The average user profile is male (92.1%), Brazilian (83.6%), and 34 years old on average. They do not receive outside incentive to participate in Wikipedia, which means that their contribution is voluntary (98.6%). They made their first edition approximately seven years ago and have made about 18,000 editions, or contributions. They occupy one or more administrative positions, such as author-reviewer, reverser, eliminator, manager, supervisor, Open-Source Ticket Request System member, or course teacher (80.4%).

**4.2. Data Preparation**

**First, we standardized the data and made a descriptive analysis of the data by verifying outliers, normality, kurtosis, and asymmetry. Two observations with indices over 3.26 (99 percent probability of being outliers) were eliminated, and the sample was reduced from 214 to 212 respondents.**

**Next, we conducted an analysis to verify normal distribution of the data for the indicators, a requirement for a major part of parametric tests. For this purpose, we assessed the asymmetry and kurtosis indices. Only four of the twenty-seven items we analyzed did not meet the criteria of having asymmetry and kurtosis values between - 2 and + 2 (Hair, Anderson, Babin, & Black, 2010). This result demonstrates that the data has univariate normal distribution.**

### 4.3. Measurement Model

The first model was tested with 27 indicators for the factors attitude, self-efficacy, altruism by identification, altruism by reciprocity, and altruism by reputation; 3 indicators for the effort construct; and the quantitative variables experience and active contributions, which present low reliability in some scales. We performed a factor analysis only on the indicators for attitude, self-efficacy, altruism by identification, altruism by reciprocity, and altruism by reputation by applying the extraction method, called the principal component analysis, and the Varimax rotation method with Kaiser normalization. Nineteen of the twenty-seven items exhibited factor loadings over 0.4 (Hair et al., 2010).

We then tested the modified model again and also complied with the suggestion to create a direct relation between experience and contribution, as this relation has been endorsed by several studies in the literature on online communities (Marchi, Giachetti, & De Gennaro, 2011; Ransbotham & Kane, 2011). The model demonstrated good fit indices, a chi-squared distribution with a significance of 0.224, a proportion of the chi-squared distribution and degrees of freedom of 1.29, CFI of 0.95, RMR of 0.065, and RMSEA of 0.032.

Regarding the validity of the latent constructs (attitude, self-efficacy, altruism by identification, altruism by reciprocity, and altruism by reputation), we noted evidence of convergent validity for attitude, self-efficacy, and altruism by identification and by reputation, since their average variance extracted values are above 0.5, according to Table 1 (Fornell & Larcker, 1981). Nevertheless, the average variance extracted for altruism by reciprocity was slightly under the value established by the literature.

Furthermore, we examined the discriminant validity of the latent constructs by comparing their square root values of the average variance extracted with their Pearson correlation coefficients. All the square root values of the average variance extracted were greater than the Pearson correlation coefficient values (r), which indicates that the model has discriminant validity from the perspective of Fornell and Larcker (1981). Table 1 displays these values.

------------------------------------

Insert Table 1 about here

-------------------------------------

The scales for attitude, self-efficacy, altruism by identification, and altruism by reputation met the levels of composite reliability and Cronbach's alpha based on the literature, with values over 0.7 (Fornell & Larcker, 1981; Nunnally, 1967). The reliability levels for altruism by reciprocity were under 0.7. Nonetheless, we decided to maintain the construct as the items of this component are related to altruism by reciprocity, which the literature indicates is a construct that is notably relevant to studying the contribution phenomenon in online communities. After performing these tests, the instrument presented general internal consistence, and it was possible to accept this factor structure.

## 4.4. Structural Model

After defining the model, we then had to verify the significance of each of the hypothesized relations. As observed, six relations were significant at the confidence interval between 0.01 and 0.05, more specifically at 0.01. Effort showed a direct association with active contributions; therefore, Hypothesis 1 was supported. Experience demonstrated a significant effect on the moderation between effort and active contributions, thus supporting Hypothesis 2a. At the same time, experience exhibited a direct association with self-efficacy and thereby confirmed Hypothesis 2b. At the level of 0.05, attitude, altruism by identification, and altruism by reputation have a direct relation with effort, confirming Hypotheses 3a, 5a, and 7a. In Figure 2, we can observe the estimates for each relation's coefficient, that is, how much each variable explains the variance of its correspondent.

-------------------------------------

Insert Figure 2 about here

-------------------------------------

Table 2 illustrates the relations among the variables in the final model and each variable's coefficient estimate, standardized coefficient, and significance in probabilistic terms. We initially proposed 13 hypotheses. Six failed to be rejected, and seven were rejected.

-----------------------------------

Insert Table 2 about here

--------------------------------

With respect to r² of the dependent variables, the total set of independent factors can explain 19 percent of the variance attained for effort and 23 percent of the variance for active contributions.

## 5. Discussion of the Results

**Based on the expectancy theory (Vroom, 1964) and the vast related literature, we observed that the more time and dedication a member invests in contributing to Wikipedia, the greater their contribution in terms of quality contributions. Similarly, experience's moderating effect on the relation between effort and active contributions was significant. Therefore, the more experience a contributor has, the more their efforts will result in relevant contributions.**

**Experience is also directly related to self-efficacy, which probably occurs because members that have collaborated longer have a better understanding of the tools and procedures necessary for contribution. Thus, they are more confident in their abilities and skills to assume such behavior, as occurs in relation to common Internet users (Eastin & Larose, 2000).**

**As for effort, the factors altruism by identification, altruism by reputation, and attitude were significant in their determination. Altruism by identification is the most relevant factor, followed by altruism by reputation and attitude. This implies that the more an individual is motivated by a sense of altruism by identification, the more effort they will exert. In other words, the more an individual feels that Wikipedia and the act of**

contributing itself are behaviors that make up who they are, the more time and dedication they will invest. Tsai and Bagozzi (2014) received similar results and discovered that a user's identification with a certain group within an online community is a factor that influences their desire to contribute.

As opposed to the positive relation with effort, the hypothesis that altruism by identification is significantly related to active contributions was not supported. A possible explanation is that altruism by identification is directly associated with the degree of proximity that the contributor feels with their peers (Fehr & Fishbacher, 2003). Therefore, contributors that have distanced themselves from the community are less motivated by altruism by identification and may have contributed significantly in the past.

This study also reveals that the more a contributor feels motivated by a sense of altruism by reputation, the more they shall perceive the effort they invest as necessary to increasing their reputation among peers. This observation corroborates the findings of an experiment that Gallus (2016) conducted on Wikipedia, which demonstrated that recent members in the community who received symbolic compensation publicly were more likely to remain active during the following weeks than those who received no compensation. The author suggests that this occurs because symbolic compensation is a mechanism that increases the individual's reputation before their peers.

The hypothesis that altruism by reputation influences active contributions was not confirmed. A possible explanation may be that members who are motivated by altruism by reputation are more preoccupied with the quality rather than the quantity of their contributions. One of the most important factors for a member to receive recognition from their peers in online communities is the perceived quality of their contributions over time

(Donath, 1999).

Concerning attitude, this study endorses the hypothesis that the more an individual believes that the results of their efforts as a Wikipedia contributor will be positive, the more favorable their attitude will be in relation to work. Consequently, they will put forth more effort, as established by the model proposed by Ajzen (2011). Although they did not examine the effort construct directly, Lin (2006) and Cho et al. (2010) discovered in their studies that attitude is a determining factor in Wikipedia collaborators' intention to contribute.

On the other hand, it is possible that members with a more favorable attitude toward Wikipedia exert more effort, but that is not directly reflected in the active contributions. One potential explanation is that attitude relative to a certain action is not static and may vary over time (Davidson & Jaccard, 1979), and because this is a transversal study, it can only assess the contributor's current attitude. For example, a member may have contributed several relevant editions for numerous years but might have a less favorable attitude today than in the past.

As for self-efficacy, it is important to note that Wikipedia collaborators may contribute not only to one task, but to a variety of tasks, including editing an existing article, creating a new article, participating in discussions about the community, and reviewing contributions by other members. Each one of these tasks presents a different complexity, which may justify why the significance of the relation between self-efficacy and the factors effort and contributions does not exist. Self-efficacy generally presents validity only when measuring specific, simple, and unique tasks (Judge, Jackson, Shaw, Scott, & Rich, 2007).

Based on the results we retrieved from this research, altruism by reciprocity does not exhibit a direct, significant relation to effort and contributions, thereby invalidating the hypotheses we proposed. This may be because in communities with strong pro-sharing norms, as appears to be the case for Wikipedia, members collaborate regardless of the perception that their efforts are being rewarded by other members (Kankanhalli et al., 2005). Accordingly, it is possible to surmise that when contributors exert effort and indeed contribute, they are more concerned with the benefits that that action will bring to the community as a whole than with receiving compensation from their peers (Chiu, Hsu, & Wang, 2006).

## 6. CONCLUSIONS

This study aims to shed light on how individual factors like attitude, self-efficacy, and altruism influence contributions and effort on Wikipedia. In addition to considering traditional indicators such as the number of members and contributions, the results indicate that it is also important for Wikipedia managers to consider indicators of effort. In other words, effort is the amount of time that a certain contributor spends on the platform and the number of citations they make in their articles, which shows greater effort in their search for information.

For research on online communities, this study's main contribution is identifying effort as a relevant factor for a better understanding of the phenomenon of voluntary contributions in online communities, more specifically, Wikipedia. Aside from discovering that the effort that members exert is directly associated with active contributions, we identified and measured factors that are positively related to effort, such as attitude, altruism by identification, and altruism by reputation. The moderating and direct role of experience also presented an interesting finding and deserves further investigation.

In turn, for the organizational literature, this study endorses the established expectancy theory upon confirming the relation between effort and performance results. It also supports the proposition by Ajzen (2011) that attitude directly influences effort. The lack of significance of the relations between individual factors and contribution requires further investigation in future studies.

We found several limitations. First, the sample is made up exclusively by members registered in Wikipedia, while anonymous users can also contribute to the community. Additionally, most of the respondents hold some administrative position in the platform. This may have possibly generated a bias in the answers that is evident when analyzing the low degree of variability in the answers regarding the attitude construct, which obtained an average above four on a scale that goes to five.

Furthermore, transversal research is unable to assess the probable changes in the individuals' characteristics and in the factors that influence their effort and contributions during the period of their participation in the platform. Finally, this study only considered registered members in the Portuguese version of the platform and chose a non-probability sampling by convenience.

For future studies, aside from individual factors, we suggest considering potential social factors that might also affect collaborators' contributions. We also recommend further exploration of the effort factor given that its relation to members' contributions has been endorsed. Finally, we propose an analysis of the individual factors' influence on active contributions and effort, which, notably, do not vary or vary slightly over time. We also suggest assessing whether personality factors, which are considerably stable over time, present some relevance to the analysis of the contribution phenomenon in Wikipedia and online communities, and if so, the extent of that influence.

**FIGURES AND TABLES**

**Figure 1 – Individual Factors that Influence Contributions on Wikipedia**

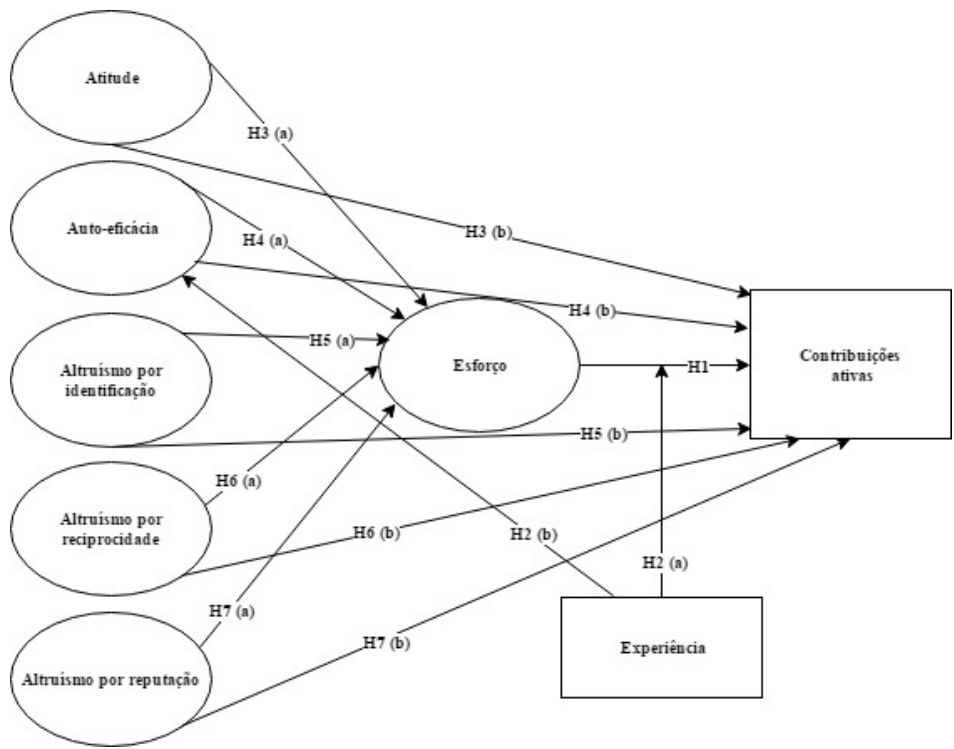

Source: The authors

**Table 1 – Average Variance Extracted (AVE), CR, Cronbach's Alpha, and Correlations.**

| Construct | Indices | | Correlations | |
|---|---|---|---|---|

| | AVE | Composite Reliability | Cronbach's Alpha | | Attitude | Self-efficacy | Altruism by Identification | Altruism by Reputation | Altruism by Reciprocity |
|---|---|---|---|---|---|---|---|---|---|
| Attitude | 0.55 | 0.83 | 0.76 | 0.74 | 1 | - | - | - | - |
| Self-efficacy | 0.61 | 0.82 | 0.67 | 0.78 | 0.07 | 1 | - | - | - |
| Altruism by Identification | 0.51 | 0.84 | 0.83 | 0.72 | .499** | .34** | 1 | - | - |
| Altruism by Reputation | 0.62 | 0.89 | 0.87 | 0.78 | .23** | .36** | .40** | 1 | - |
| Altruism by Reciprocity | 0.45 | 0.62 | 0.40 | 0.67 | .31** | .22** | .35** | .50** | 1 |

Note: **p < 0.01

*p < 0.05

**Figure 2 – Structural Equation Results**

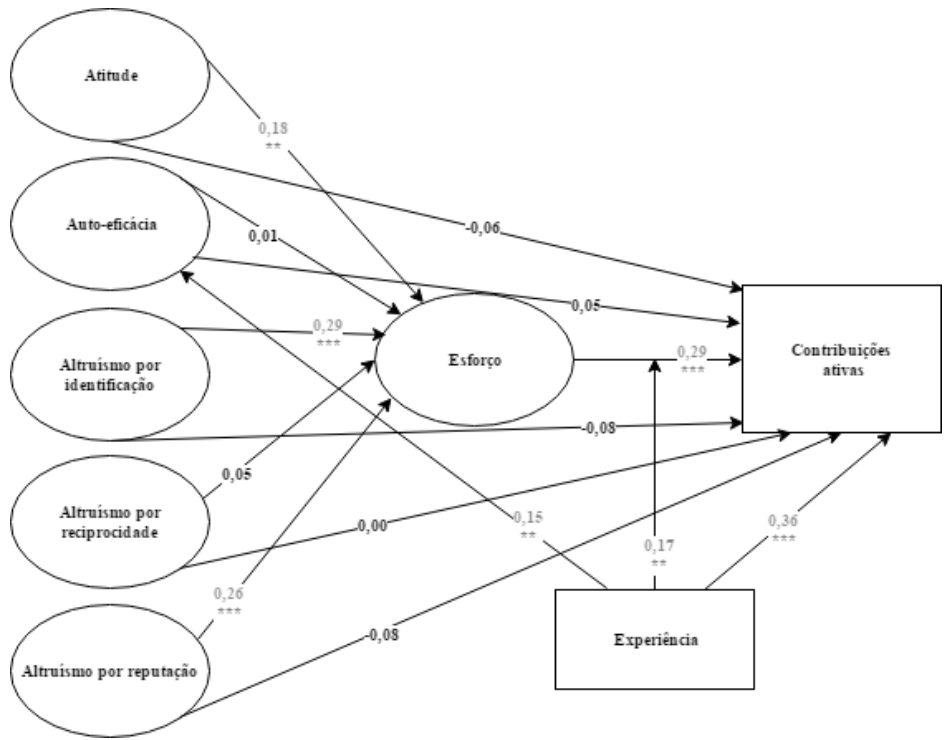

Note: *** Significant at 1%

    ** Significant at 5%

Source: The authors

**Table 2 – Relations, Estimates, Standardized Coefficients, and Hypotheses Tests**

| Relationships | | Estimative | Z | P | Hypothesis | Result |
|---|---|---|---|---|---|---|
| Effort | Active Contributions | 0.29 | 4.31 | *** | H1 | Failed to be rejected |
| Effort x experience | Active Contributions | 0.17 | 2.79 | .01** | H2a | Failed to be rejected |
| Experience | Self-efficacy | 0.15 | 2.20 | .03** | H2b | Failed to be rejected |
| Experience | Active Contributions | 0.36 | 5.93 | *** | | |
| Attitude | Effort | 0.18 | 2.97 | .00** | H3a | Failed to be rejected |
| Attitude | Active Contributions | - 0.06 | - 0.95 | 0.34 | H3b | Rejected |
| Self-efficacy | Effort | 0.01 | 0.10 | 0.92 | H4a | Rejected |
| Self-efficacy | Active Contributions | 0.05 | 0.88 | 0.38 | H4b | Rejected |
| Altruism by Identification | Effort | 0.29 | 4.69 | *** | H5a | Failed to be rejected |
| Altruism by Identification | Active Contributions | - 0.08 | - 1.83 | 0.07 | H5b | Rejected |
| Altruism by Reciprocity | Effort | 0.05 | 0.78 | 0.44 | H6a | Rejected |
| Altruism by Reciprocity | Active Contributions | 0.00 | 0.03 | 0.98 | H6b | Rejected |
| Altruism by Reputation | Effort | 0.26 | 4.23 | *** | H7a | Failed to be rejected |
| Altruism by Reputation | Active Contributions | - 0.08 | - 1.24 | 0.22 | H7b | Rejected |